\documentclass[fleqn,11pt]{article}

\usepackage{amsmath,amsfonts,amssymb}
\usepackage{color}

\usepackage{geometry}
\newtheorem{theorem}{Theorem}
\newtheorem{lemma}[theorem]{Lemma}

\newcommand{\R}{\mathbb{R}}

\newcommand{\N}{\mathbb{N}}

\newcommand{\cqfd}
{%
\mbox{}%
\nolinebreak%
\hfill%
\rule{2mm}{2mm}%
\newline
\newline
}

\title{
Stationary solutions to the two-dimensional Broadwell model.}
\author{Leif Arkeryd and Anne Nouri}
\date{}

\begin{document}
\maketitle
\hspace{1cm}\\
{\bf Abstract} \hspace{1cm}\\
Existence of renormalized solutions to the two-dimensional Broadwell model with given indata in $L^1$ is proven. Averaging techniques from the continuous velocity case being  unavailable when  the velocities are discrete, the approach is based on direct $L^1$-compactness arguments using the Kolmogorov-Riesz theorem.
\footnotetext[1]{2010 Mathematics Subject Classification 82C22, 82C40.}
\footnotetext[2]{Key words; Broadwell equation, discrete velocity Boltzmann equation, existence theory.}
\footnotetext[3]{Leif Arkeryd, Mathematical Sciences, Goteborg, Sweden.}
\footnotetext[4]{Anne Nouri, Aix-Marseille University, CNRS, Centrale Marseille, I2M, Marseille, France.}
%
%
%
%
\section {Introduction.}
 \setcounter{equation}{0}
 \setcounter{theorem}{0}
The two-dimensional stationary Broadwell model in a square is
\begin{align}
&\partial_xF_1= F_3F_4-F_1F_2,&F_1(0,\cdot )= f_{b1},\nonumber \\
&-\partial_xF_2= F_3F_4-F_1F_2,&F_2(1,\cdot )= f_{b2},\nonumber \\
&\partial_yF_3= F_1F_2-F_3F_4,&F_3(\cdot ,0)= f_{b3},\nonumber\\
&-\partial_yF_4= F_1F_2-F_3F_4,&F_4(\cdot ,1)= f_{b4},\label{eq-broadwell}
\end{align}
with unknown $(F_i)_{1\leq i\leq 4}$ defined on $[0,1]^2$, and given $(f_{bi})_{1\leq i\leq 4}$ defined on $[0,1]$. It is a four velocity model for the Boltzmann equation, with $F_i(x,y)= f(x,y,v_i)$, 
\begin{align*}
&v_1= (1,0),\quad v_2= (-1,0),\quad v_3= (0,1),\quad v_4= (0,-1).
\end{align*}
The boundary value problem \eqref{eq-broadwell} is considered in $L^1$ in one of the following equivalent
forms, \\
 the exponential multiplier form:
\begin{align}\label{exponential-form}
&F_{1}(x,y)= f_{b1}(y)e^{-\int_0^x F_2(s,y)ds}+\int_0^x (F_3F_4)(s,y)e^{-\int_s^x F_2(\tau)d\tau}ds ,\quad \text{a.a.  }(x,y)\in [0,1]^2,
\end{align}
and analogous equations for $F_i$, $2\leq i\leq 4$,\\
 the mild form:
\begin{align}\label{mild-form}
&F_1(x,y)= f_{b1}(y)+\int_0^x (F_3F_4-F_1F_2)(s,y)ds,\quad \text{a.a.  }(x,y)\in [0,1]^2,
\end{align}
and analogous equations for $F_i$, $2\leq i\leq 4$,\\
the renormalized form:
\begin{align}\label{renormalized-form}
\partial_x\ln(1+F_1)= \frac{F_3F_4-F_1F_2}{1+F_1},&F_1(0,\cdot )= f_{b1},
\end{align}
in  the sense of distributions, and analogous equations for $F_i$, $2\leq i\leq 4$.\\
%
%
%
The main result of the paper is the following.
\begin{theorem}\label{global-existence-Broadwell}
\hspace*{1.in}\\
Given a non-negative boundary value $f_{b}$ with finite mass and entropy, there exists a stationary non-negative renormalized solution in $L^1$ with finite entropy-dissipation to the Broadwell model \eqref{eq-broadwell}.
\end{theorem}
%
%
Most mathematical results for discrete velocity models of the Boltzmann equation have been performed in one space dimension. An overview is given in \cite{IP}. 
In two dimensions, special classes of solutions are given in \cite{B} 
\cite{BT},
and  \cite{Ily}. \cite{B} contains a detailed study of the stationary Broadwell equation in a rectangle with comparison to a 
Carleman-like system, and a discussion of (in)compressibility aspects.\\
\\
The existence of continuous solutions to the two-dimens{ional stationary Broadwell model with continuous boundary data for a rectangle, is proven in \cite{CIS}. That proof starts by solving the problem with a given gain term, and uses the compactness of the corresponding twice iterated  solution operator to conclude by Schaeffer's fixed point theorem. \\
 The present paper on the Broadwell model is set in a context of physically natural quantities. Mass and entropy flow at the boundary are given, and the solutions obtained, have finite mass and finite entropy dissipation. Averaging techniques from the continuous velocity case \cite{DPL} being unavailable, a direct compactness approach is used, 
based on the Kolmogorov-Riesz theorem.\\
The plan of the paper is the following. An  approximation procedure  for the construction of solutions 
to (1.1) is introduced in Section \ref{approximations}. The passage to the limita in the approximations is performed in Section 3.
Here a compactness property of the  approximated gain terms in mild form is carried over to the corresponding solutions themselves, using  a particular sequence of successive  alternating approximations and the Kolmogorov-Riesz theorem \cite{K}, \cite{R}. The approach also holds for domains which are strictly convex with $C^1$ boundary.\\
A common approach to existence for stationary Boltzmann like equations is based on the regularizing properties of the gain term. In the continuous velocity case an averaging propery is available  to keep this study of the gain term within a weak $L^1$ frame as in \cite{AN1}. However, 
in the discrete velocity case, averaging is not available. Instead  strong convergence of an approximating sequence is  here directly proved  from the regularizing  properties for the gain term (cf Lemma 3.5 below). But the technique in that proof is restricted  to two dimensional velocities, whereas the averaging technique in the continuous velocity case is dimension independent.

\section{Approximations.}
\label{approximations}
 \setcounter{equation}{0}
 \setcounter{theorem}{0}
%
%
Denote by $L^1_+([0,1]^2)$ the set of non negative integrable functions on $[0,1]^2$, and by $a\wedge b$ the minimum of two real numbers $a$ and $b$. Approximations to \eqref{eq-broadwell} to be used in the proof of Theorem 1, are introduced in the following lemma.
\begin{lemma}\label{approximations-broadwell}
For any $k\in \N ^*$, there exists a solution $F^k\in \big( L^1_+([0,1]^2)\big) ^4$ to
\begin{align}
&\partial _xF^k_1= \frac{F^k_3}{1+\frac{F^k_3}{k}}\frac{F^k_4}{1+\frac{F^k_4}{k}}-\frac{F^k_1}{1+\frac{F^k_1}{k}}\frac{F^k_2}{1+\frac{F^k_2}{k}}\hspace*{0.03in},\label{Fk-1}\\
&-\partial _xF^k_2= \frac{F^k_3}{1+\frac{F^k_3}{k}}\frac{F^k_4}{1+\frac{F^k_4}{k}}-\frac{F^k_1}{1+\frac{F^k_1}{k}}\frac{F^k_2}{1+\frac{F^k_2}{k}}\hspace*{0.03in},\label{Fk-2}\\
&\partial _yF^k_3= \frac{F^k_1}{1+\frac{F^k_1}{k}}\frac{F^k_2}{1+\frac{F^k_2}{k}}-\frac{F^k_3}{1+\frac{F^k_3}{k}}\frac{F^k_4}{1+\frac{F^k_4}{k}}\hspace*{0.03in},\label{Fk-3}\\
&-\partial _yF^k_4= \frac{F^k_1}{1+\frac{F^k_1}{k}}\frac{F^k_2}{1+\frac{F^k_2}{k}}-\frac{F^k_3}{1+\frac{F^k_3}{k}}\frac{F^k_4}{1+\frac{F^k_4}{k}},\quad (x,y)\in [0,1]^2,\label{Fk-4}\\
&F^k_1(0,y)= f_{b1}(y)\wedge \frac{k}{2},\quad F^k_2(1,y)= f_{b2}(y)\wedge \frac{k}{2},\quad y\in [0,1],\label{bcFk-1-Fk-2} \\
&F^k_3(x,0)= f_{b3}(x)\wedge \frac{k}{2},\quad F^k_4(x,1)= f_{b4}(x)\wedge \frac{k}{2},\quad x\in [0,1].\label{bcFk-3-Fk-4}
\end{align} 
\end{lemma}
\underline{Proof of Lemma \ref{approximations-broadwell}.}\\
The sequence of approximations $(F^k)_{k\in \N ^*}$ is obtained in the limit of a further approximation with damping terms $\alpha F_j$ and convolutions in the collision operator.\\
\hspace*{1.in}\\ 
\underline{Step I. Approximations with damping and convolutions.}\\
Take  $\alpha >0$ and set
\begin{align}\label{df-c-alpha}
&c_\alpha = \frac{1}{\alpha }\int _0^1\sum _{i=1}^4f_{bi}(u)du,\quad K_\alpha=\{f\in \big( L_+^1([0,1]^2)\big) ^4;\sum _{i=1}^4\int f_i(x,y)dxdy\leq c_\alpha\} .
\end{align} 
Let $\mu_\alpha$ be a smooth mollifier in $(x,y)$ with support in the ball centered at the origin of radius $\alpha $. 
Let $\mathcal{T}$ be  the map defined on $K_\alpha $ by $\mathcal{T}(f)=F$, where $F= (F_i)_{1\leq i\leq 4}$ is the solution of
\begin{align}
&\alpha F_1+\partial _xF_1= \frac{F_3}{1+\frac{F_3}{k}}\frac{f_4\ast \mu_\alpha }{1+\frac{f_4\ast \mu_\alpha }{k}}-\frac{F_1}{1+\frac{F_1}{k}}\frac{f_2\ast \mu_\alpha }{1+\frac{f_2\ast \mu_\alpha }{k}}\hspace*{0.03in},\label{df-F1}\\
&\alpha F_2-\partial _xF_2= \frac{f_3\ast \mu_\alpha }{1+\frac{f_3\ast \mu_\alpha }{k}}\frac{F_4}{1+\frac{F_4}{k}}-\frac{f_1\ast \mu_\alpha }{1+\frac{f_1\ast \mu_\alpha }{k}}\frac{F_2}{1+\frac{F_2}{k}}\hspace*{0.03in},\label{df-F2}\\
&\alpha F_3+\partial _yF_3= \frac{F_1}{1+\frac{F_1}{k}}\frac{f_2\ast \mu_\alpha }{1+\frac{f_2\ast \mu_\alpha }{k}}-\frac{F_3}{1+\frac{F_3}{k}}\frac{f_4\ast \mu_\alpha }{1+\frac{f_4\ast \mu_\alpha }{k}}\hspace*{0.03in},\label{df-F3}\\
& \alpha F_4-\partial _yF_4= \frac{f_1\ast \mu_\alpha }{1+\frac{f_1\ast \mu_\alpha }{k}}\frac{F_2}{1+\frac{F_2}{k}}-\frac{f_3\ast \mu_\alpha }{1+\frac{f_3\ast \mu_\alpha }{k}}\frac{F_4}{1+\frac{F_4}{k}}\hspace*{0.03in},\quad (x,y)\in [0,1]^2,\label{df-F4}\\
&F_1(0,y)= f_{b1}(y)\wedge \frac{k}{2},\quad F_2(1,y)= f_{b2}(y)\wedge \frac{k}{2},\quad y\in [0,1],\label{bcF1-F2} \\
&F_3(x,0)= f_{b3}(x)\wedge \frac{k}{2},\quad F_4(x,1)= f_{b4}(x)\wedge \frac{k}{2},\quad x\in [0,1].\label{bcF3-F4}
\end{align}
$F= \mathcal{T}(f)$ is obtained as the limit in $L^1([0,1]^2)$ of the sequence $(F^n)_{n\in \N }$ defined by $F^0= 0$ and 
\begin{align*}
&\alpha F_1^{n+1}+\partial _xF_1^{n+1}= \frac{F_3^n}{1+\frac{F_3^n}{k}}\frac{f_4\ast \mu_\alpha }{1+\frac{f_4\ast \mu_\alpha }{k}}-\frac{F_1^{n+1}}{1+\frac{F_1^n}{k}}\frac{f_2\ast \mu_\alpha }{1+\frac{f_2\ast \mu_\alpha }{k}},\\
& \alpha F_2^{n+1}-\partial _xF_2^{n+1}= \frac{f_3\ast \mu_\alpha }{1+\frac{f_3\ast \mu_\alpha }{k}}\frac{F_4^n}{1+\frac{F_4^n}{k}}-\frac{f_1\ast \mu_\alpha }{1+\frac{f_1\ast \mu_\alpha }{k}}\frac{F_2^{n+1}}{1+\frac{F_2^n}{k}},\\
&\alpha F_3^{n+1}+\partial _yF_3^{n+1}= \frac{F_1^n}{1+\frac{F_1^n}{k}}\frac{f_2\ast \mu_\alpha }{1+\frac{f_2\ast \mu_\alpha }{k}}-\frac{F_3^{n+1}}{1+\frac{F_3^n}{k}}\frac{f_4\ast \mu_\alpha }{1+\frac{f_4\ast \mu_\alpha }{k}},\\
& \alpha F_4^{n+1}-\partial _yF_4^{n+1}= \frac{f_1\ast \mu_\alpha }{1+\frac{f_1\ast \mu_\alpha }{k}}\frac{F_2^n}{1+\frac{F_2^n}{k}}-\frac{f_3\ast \mu_\alpha }{1+\frac{f_3\ast \mu_\alpha }{k}}\frac{F_4^{n+1}}{1+\frac{F_4^n}{k}},\\
&F_1^{n+1}(0,y)= f_{b1}(y)\wedge \frac{k}{2},\quad F_2^{n+1}(1,y)= f_{b2}(y)\wedge \frac{k}{2},\quad y\in [0,1],\\
&F_3^{n+1}(x,0)= f_{b3}(x)\wedge \frac{k}{2},\quad F_4^{n+1}(x,1)= f_{b4}(x)\wedge \frac{k}{2},\quad x\in [0,1],\quad n\in \N .
\end{align*}
The sequence $(F^n)_{n\in \N }$ is monotone. Indeed, $F_i^0\leq F_i^1$, $1\leq i\leq 4$ by the exponential form of $F_i^1$. Moreover, assume $F_i^n\leq F_i^{n+1}$, $1\leq i\leq 4$. It follows from the exponential form 
%
%
that $F_1^{n+1}-F_1^{n+2}\leq 0$. The inequalities $F_i^{n+1}-F_i^{n+2}\leq 0$, $2\leq i\leq 4$ can be reached in a similar way. Moreover,
\begin{align*}
&\alpha \sum _{i=1}^4F_i^{n+1}+\partial _x(F_1^{n+1}-F_2^{n+1})+\partial _y(F_3^{n+1}-F_4^{n+1})\\
&= \frac{f_1\ast \mu_\alpha }{1+\frac{f_1\ast \mu_\alpha }{k}}\frac{F_2^n-F_2^{n+1}}{1+\frac{F_2^n}{k}}+\frac{f_2\ast \mu_\alpha }{1+\frac{f_2\ast \mu_\alpha }{k}}\frac{F_1^n-F_1^{n+1}}{1+\frac{F_1^n}{k}}+\frac{f_3\ast \mu_\alpha }{1+\frac{f_3\ast \mu_\alpha }{k}}\frac{F_4^n-F_4^{n+1}}{1+\frac{F_4^n}{k}}+\frac{f_4\ast \mu_\alpha }{1+\frac{f_4\ast \mu_\alpha }{k}}\frac{F_3^n-F_3^{n+1}}{1+\frac{F_3^n}{k}}\\
&\leq 0,
\end{align*}
so that
\begin{align}\label{total-mass-Fn-i}
&\sum _{i=1}^4\int F_i^{n+1}(x,y)dxdy\leq c_\alpha .
\end{align}
By the monotone convergence theorem, $(F^n)_{n\in \N }$ converges in $L^1([0,1]^2)$ to some solution $F$ of (\ref{df-F1})-(\ref{bcF3-F4}). The solution of (\ref{df-F1})-(\ref{bcF3-F4}) is unique in the set of non negative functions. Indeed, let $G=(G_i)_{1\leq i\leq 4}$ be a solution of \eqref{df-F1}-\eqref{bcF3-F4} with $G_i\geq 0$, $1\leq i\leq 4$. Let us prove by induction that 
\begin{align}\label{recurrence1}
&\forall n\in \N , \quad F_i^n\leq G_i,\quad 1\leq i\leq 4.
\end{align}
\eqref{recurrence1} holds for $n= 0$, since $G_i\geq 0$, $1\leq i\leq 4$. Assume \eqref{recurrence1} holds for $n$. Using the exponential form of $F_1^{n+1}$ implies 
\begin{align*}
F_1^{n+1}(x,y)&= (f_{b1}(y)\wedge \frac{k}{2})e^{-\alpha x-\int _0^x\frac{f_2\ast \mu_\alpha }{(1+\frac{F^n_1}{k})(1+\frac{f_2\ast \mu_\alpha }{k})}(X,y)dX}\\
&+\int _0^x\frac{F_3^n}{1+\frac{F_3^n}{k}}\frac{f_4\ast \mu_\alpha }{1+\frac{f_4\ast \mu_\alpha }{k}}(X,y)e^{-\alpha (x-X)-\int _X^x\frac{f_2\ast \mu_\alpha }{(1+\frac{F^n_1}{k})(1+\frac{f_2\ast \mu_\alpha }{k})}(r,y)dr}dX\\
&\leq (f_{b1}(y)\wedge \frac{k}{2})e^{-\alpha x-\int _0^x\frac{f_2\ast \mu_\alpha }{(1+\frac{G_1}{k})(1+\frac{f_2\ast \mu_\alpha }{k})}(X,y)dX}\\
&+\int _0^x\frac{G_3}{1+\frac{G_3}{k}}\frac{f_4\ast \mu_\alpha }{1+\frac{f_4\ast \mu_\alpha }{k}}(X,y)e^{-\alpha (x-X)-\int _X^x\frac{f_2\ast \mu_\alpha }{(1+\frac{G_1}{k})(1+\frac{f_2\ast \mu_\alpha }{k})}(r,y)dr}dX\\
&= G_1(x,y),\quad (x,y)\in [0,1]^2 .
\end{align*}
The same argument can be applied to prove that $F_i^{n+1}\leq G_i$, $2\leq i\leq 4$. Consequently,
\begin{align}\label{comparaison-F-G}
&F_i\leq G_i,\quad 1\leq i\leq 4.
\end{align}
Moreover, substracting the partial differential equations satisfied by $G_i$ from the partial differential equations satisfied by $F_i$, $1\leq i\leq 4$, and integrating the resulting equation on $[0,1]^2$, it results
\begin{align}\label{total-mass-F-i}
\alpha \sum _{i=1}^4\int (G_i-F_i)(x,y)dxdy&+\int _0^1\big( (G_1-F_1)(1,y)+(G_2-F_2)(0,y)\big) dy\nonumber \\
&+\int _0^1\big( (G_3-F_3)(x,1)+(G_4-F_4)(x,0)\big) dx= 0.
\end{align}
It results from \eqref{comparaison-F-G}-\eqref{total-mass-F-i} that $G= F$.\\
The map $\mathcal{T}$ is continuous in the $L^1$-norm topology (cf [1] pages 124-5). Namely, let a sequence $(f_l)_{l\in \N }$ in $K_\alpha$ converge in $L^1([0,1]^2)$ to $f\in K_\alpha$. Set $F_l=T(f_l)$. Because of the uniqueness of the solution to (\ref{df-F1})-(\ref{bcF3-F4}), it is enough to prove that there is a subsequence of  $(F_l)$ converging to $F=T(f)$. Now there is a subsequence of $(f_l)$, still denoted $(f_l)$, such that decreasingly (resp. increasingly) $(G_l)= (\sup _{m\geq l}f_m)$ (resp. $(g_l)= (\inf _{m\geq l}f_m)$)  converges to $f$ in $L^1$. Let $(S_l)$ (resp. $(s_l)$) be the sequence of solutions to
\begin{align}\label{df-Sl}
&\alpha S_{l1}+\partial _xS_{l1}= \frac{S_{l3}}{1+\frac{S_{l3}}{k}}\frac{G_{l4}\ast \mu_\alpha }{1+\frac{G_{l4}\ast \mu _\alpha }{k}}-\frac{S_{l1}}{1+\frac{S_{l1}}{k}}\frac{g_{l2}\ast \mu_\alpha }{1+\frac{g_{l2}\ast \mu _\alpha }{k}}\hspace*{0.03in},\\
& \alpha S_{l2}-\partial _xS_{l2}= \frac{G_{l3}\ast \mu_\alpha }{1+\frac{G_{l3}\ast \mu _\alpha }{k}}\frac{S_{l4}}{1+\frac{S_{l4}}{k}}-\frac{g_{l1}\ast \mu_\alpha }{1+\frac{g_{l1}\ast \mu _\alpha }{k}}\frac{S_{l2}}{1+\frac{S_{l2}}{k}}\hspace*{0.03in},\\
&\alpha S_{l3}+\partial _yS_{l3}= \frac{S_{l1}}{1+\frac{S_{l1}}{k}}\frac{G_{l2}\ast \mu_\alpha }{1+\frac{G_{l2}\ast \mu _\alpha }{k}}-\frac{S_{l3}}{1+\frac{S_{l3}}{k}}\frac{g_{l4}\ast \mu_\alpha }{1+\frac{g_{l4}\ast \mu _\alpha }{k}}\hspace*{0.03in},\\
& \alpha S_{l4}-\partial _yS_{l4}= \frac{G_{l1}\ast \mu_\alpha }{1+\frac{G_{l1}\ast \mu _\alpha }{k}}\frac{S_{l2}}{1+\frac{S_{l2}}{k}}-\frac{g_{l3}\ast \mu_\alpha }{1+\frac{g_{l3}\ast \mu _\alpha }{k}}\frac{S_{l4}}{1+\frac{S_{l4}}{k}}\hspace*{0.03in},\\
&S_{l1}(0,y)= f_{b1}(y)\wedge \frac{k}{2},\quad S_{l2}(1,y)= f_{b2}(y)\wedge \frac{k}{2},\quad y\in [0,1],\\
&S_{l3}(x,0)= f_{b3}(x)\wedge \frac{k}{2},\quad S_{l4}(x,1)= f_{b4}(x)\wedge \frac{k}{2},\quad x\in [0,1],
\end{align}
(resp.
\begin{align}\label{df-Sl}
&\alpha s_{l1}+\partial _xs_{l1}= \frac{s_{l3}}{1+\frac{s_{l3}}{k}}\frac{g_{l4}\ast \mu_\alpha }{1+\frac{g_{l4}\ast \mu _\alpha }{k}}-\frac{s_{l1}}{1+\frac{s_{l1}}{k}}\frac{G_{l2}\ast \mu_\alpha }{1+\frac{G_{l2}\ast \mu _\alpha }{k}}\hspace*{0.03in},\\
& \alpha s_{l2}-\partial _xs_{l2}= \frac{g_{l3}\ast \mu_\alpha }{1+\frac{g_{l3}\ast \mu _\alpha }{k}}\frac{s_{l4}}{1+\frac{s_{l4}}{k}}-\frac{G_{l1}\ast \mu_\alpha }{1+\frac{G_{l1}\ast \mu _\alpha }{k}}\frac{s_{l2}}{1+\frac{s_{l2}}{k}}\hspace*{0.03in},\\
&\alpha s_{l3}+\partial _ys_{l3}= \frac{s_{l1}}{1+\frac{s_{l1}}{k}}\frac{g_{l2}\ast \mu_\alpha }{1+\frac{g_{l2}\ast \mu _\alpha }{k}}-\frac{s_{l3}}{1+\frac{s_{l3}}{k}}\frac{G_{l4}\ast \mu_\alpha }{1+\frac{G_{l4}\ast \mu _\alpha }{k}}\hspace*{0.03in},\\
& \alpha s_{l4}-\partial _ys_{l4}= \frac{g_{l1}\ast \mu_\alpha }{1+\frac{g_{l1}\ast \mu _\alpha }{k}}\frac{s_{l2}}{1+\frac{s_{l2}}{k}}-\frac{G_{l3}\ast \mu_\alpha }{1+\frac{G_{l3}\ast \mu _\alpha }{k}}\frac{s_{l4}}{1+\frac{s_{l4}}{k}}\hspace*{0.03in},\\
&s_{l1}(0,y)= f_{b1}(y)\wedge \frac{k}{2},\quad s_{l2}(1,y)= f_{b2}(y)\wedge \frac{k}{2},\quad y\in [0,1],\\
&s_{l3}(x,0)= f_{b3}(x)\wedge \frac{k}{2},\quad s_{l4}(x,1)= f_{b4}(x)\wedge \frac{k}{2},\quad x\in [0,1]).
\end{align}
$(S_l)$ is a non-increasing sequence, since that holds for the successive iterates defining the sequence. Then $(S_l)$ decreasingly converges in $L^1$ to some $S$. Similarly $(s_l)$ increasingly converges in $L^1$ to some $s$. The limits $S$ and $s$ satisfy \eqref{df-F1}-\eqref{bcF3-F4}. It follows by uniqueness that $s= F= S$, hence that $(F_l)$ converges in $L^1$ to $F$.\\
The map $\mathcal{T}$ is also compact in the $L^1$-norm topology. Indeed, let $(f_l)_{l\in \N }$ be a sequence in $K_\alpha$ and $(F_l)_{l\in \N}= (\mathcal{T}(f_l))_{l\in \N }$. For any $\lvert h\rvert <1$, denote by $G_{l1}(x,y)= F_{l1}(x,y+h)-F_{l1}(x,y)$ and 
\begin{align*}
H_{l1}(x,y)= &\frac{F_{l3}}{1+\frac{F_{l3}}{k}}\frac{f_{l4}\ast \mu _\alpha }{1+\frac{f_{l4}\ast \mu _\alpha }{k}}(x,y+h)-\frac{F_{l3}}{1+\frac{F_{l3}}{k}}\frac{f_{l4}\ast \mu _\alpha }{1+\frac{f_{l4}\ast \mu _\alpha }{k}}(x,y)\\
&-\frac{F_{l1}}{1+\frac{F_{l1}}{k}}(x,y+h)\Big( \frac{f_{l2}\ast \mu _\alpha }{1+\frac{f_{l2}\ast \mu _\alpha }{k}}(x,y+h)-\frac{f_{l2}\ast \mu _\alpha }{1+\frac{f_{l2}\ast \mu _\alpha }{k}}(x,y)\Big) 
\end{align*}
 They satisfy
\begin{align*}
&\big( \alpha +\frac{f_{l2}\ast \mu _\alpha }{1+\frac{f_{l2}\ast \mu _\alpha }{k}}\big) G_{l1}+\partial _xG_{l1}=  H_{l1},\quad G_{l1}(0,\cdot )= 0,
\end{align*}
so that
\begin{align*}
&G_{l1}(x,y)= \int _0^xH_{l1}(X,y)e^{-\alpha (x-X)-\int _X^x\frac{f_{l2}\ast \mu _\alpha }{1+\frac{f_{l2}\ast \mu _\alpha }{k}}(u,y)du}dX,\quad (x,y)\in [0,1]^2.
\end{align*}
The boundedness by $k^2$ of the integrands in the r.h.s. of (\ref{df-F1}) and (\ref{df-F3}) induces uniform $L^1$-equicontinuity of $(F_{l1})_{l\in \N }$ (resp. $(F_{l3})_{l\in \N }$) w.r.t. the $x$ (resp. $y$) variable. Together with the $L^1$-compactness of $(f_l\ast \mu _\alpha )_{l\in \N }$, this implies uniform $L^1$-equicontinuity w.r.t. the $y$ variable of $(H_{l1})_{l\in \N }$, then of $(F_{l1})_{l\in \N }$. This proves the $L^1$ compactness of $(F_{l1})_{l\in \N}$. The $L^1$ compactness of $(F_{li})_{l\in \N}$, $2\leq i\leq 4$ can be proven similarly.\\
Hence by the Schauder fixed point theorem there is a fixed point $\mathcal{T}(F)=F$, i.e. a solution $F$ to
\begin{align}
&\alpha F_1+\partial _xF_1= \frac{F_3}{1+\frac{F_3}{k}}\frac{F_4\ast \mu_\alpha }{1+\frac{F_4\ast \mu_\alpha }{k}}-\frac{F_1}{1+\frac{F_1}{k}}\frac{F_2\ast \mu_\alpha }{1+\frac{F_2\ast \mu_\alpha }{k}}\hspace*{0.03in},\label{F1-alpha}\\
&\alpha F_2-\partial _xF_2= \frac{F_3\ast \mu_\alpha }{1+\frac{F_3\ast \mu_\alpha }{k}}\frac{F_4}{1+\frac{F_4}{k}}-\frac{F_1\ast \mu_\alpha }{1+\frac{F_1\ast \mu_\alpha }{k}}\frac{F_2}{1+\frac{F_2}{k}}\hspace*{0.03in},\label{F2-alpha}\\
&\alpha F_3+\partial _yF_3= \frac{F_1}{1+\frac{F_1}{k}}\frac{F_2\ast \mu_\alpha }{1+\frac{F_2\ast \mu_\alpha }{k}}-\frac{F_3}{1+\frac{F_3}{k}}\frac{F_4\ast \mu_\alpha }{1+\frac{F_4\ast \mu_\alpha }{k}}\hspace*{0.03in},\label{F3-alpha}\\
& \alpha F_4-\partial _yF_4= \frac{F_1\ast \mu_\alpha }{1+\frac{F_1\ast \mu_\alpha }{k}}\frac{F_2}{1+\frac{F_2}{k}}-\frac{F_3\ast \mu_\alpha }{1+\frac{F_3\ast \mu_\alpha }{k}}\frac{F_4}{1+\frac{F_4}{k}}\hspace*{0.03in},\quad (x,y)\in [0,1]^2\label{F4-alpha}\\
&F_1(0,y)= f_{b1}(y)\wedge \frac{k}{2},\quad F_2(1,y)= f_{b2}(y)\wedge \frac{k}{2},\quad y\in [0,1],\label{bcF1-F2-alpha} \\
&F_3(x,0)= f_{b3}(x)\wedge \frac{k}{2},\quad F_4(x,1)= f_{b4}(x)\wedge \frac{k}{2},\quad x\in [0,1].\label{bcF3-F4-alpha}
\end{align} 
\\
\underline{Step II. Removal of the damping and the convolutions in (\ref{F1-alpha})-(\ref{bcF3-F4-alpha})}.\\
\\Let $k>1$ be fixed. Denote by $F^\alpha $ the solution to \eqref{F1-alpha}-\eqref{bcF3-F4-alpha} defined in Step I. Each component of $F^\alpha $ being bounded by a multiple of $k^2$, $(F^\alpha )_{\alpha \in ]0,1[}$ is weakly compact in $L^1([0,1]^2)$. Denote by $F^k$ a limit of a subsequence for the weak topology of $L^1([0,1]^2)$. Let us prove that the convergence is strong in $L^1([0,1]^2)$. Consider the approximation scheme $(f^{\alpha ,l}_1,f^{\alpha ,l}_2)_{l\in \N }$ of $(F^\alpha _1,F^\alpha _2)_{\alpha \in ]0,1[}$,

\begin{align}\label{approx-scheme-delta-1}
&f^{\alpha ,0}_1=f^{\alpha ,0}_2= 0,\nonumber \\
&\alpha f^{\alpha ,l+1}_1+\partial _xf^{\alpha ,l+1}_1= \frac{F^\alpha _3}{1+\frac{F^\alpha _3}{k}}\frac{F^\alpha _4\ast \mu_\alpha }{1+\frac{F^\alpha _4\ast \mu_\alpha }{k}}-\frac{f^{\alpha ,l+1}_1}{1+\frac{f^{\alpha ,l+1}_1}{k}}\frac{f^{\alpha ,l}_2\ast \mu _\alpha }{1+\frac{f^{\alpha ,l}_2\ast \mu _\alpha }{k}},\quad f^{\alpha ,l+1}_1(0,y)= f_{b1}(y)\wedge \frac{k}{2},\nonumber \\
&\alpha f^{\alpha ,l+1}_2-\partial _xf^{\alpha ,l+1}_2= \frac{F^\alpha _3}{1+\frac{F^\alpha _3}{k}}\frac{F^\alpha _4\ast \mu_\alpha }{1+\frac{F^\alpha _4\ast \mu_\alpha }{k}}-\frac{f^{\alpha ,l}_1\ast \mu _\alpha }{1+\frac{f^{\alpha ,l}_1\ast \mu _\alpha }{k}}\frac{f^{\alpha ,l+1}_2}{1+\frac{f^{\alpha ,l+1}_2}{k}},\quad f^{\alpha ,l+1}_2(1,y)= f_{b2}(y)\wedge \frac{k}{2},\nonumber \\
&\hspace*{5.in}l\in \N .
\end{align}
By induction on $l$ it holds that
\begin{align}\label{order-f-alpha-l}
&f_1^{\alpha ,2l}\leq f_1^{\alpha ,2l+2}\leq F_1^\alpha \leq f_1^{\alpha ,2l+3}\leq f_1^{\alpha ,2l+1},\nonumber \\
&f_2^{\alpha ,2l}\leq f_2^{\alpha ,2l+2}\leq F_2^\alpha \leq f_2^{\alpha ,2l+3}\leq f_2^{\alpha ,2l+1},\quad \alpha \in ]0,1[ ,\quad l\in \N .
\end{align}
For every $l\in \N$, $(f^{\alpha ,l}_1)_{\alpha \in ]0,1[}$ (resp. $(f^{\alpha ,l}_2)_{\alpha \in ]0,1[}$) is translationnaly equicontinuous in the $x$-direction, since all integrands in its exponential form are bounded. It is translationnaly $L^1$-equicontinuous in the $y$-direction by induction on $l$. Indeed, it is so for $(F^\alpha _3)$ (resp. $(F^\alpha _4)$) since $\partial _y(e^{\alpha y}F^\alpha _3)$ ( resp. $\partial _y(e^{\alpha y}F^\alpha _4)$) is bounded by $ek^2$, and $(\frac{F^\alpha _i}{1+\frac{F^\alpha _i}{k}})_{\alpha \in ]0,1[}$, $i\in \{ 3,4\} $, is bounded by $k$. Consequently, it is so for $(\frac{F^\alpha _3}{1+\frac{F^\alpha _3}{k}}\frac{F^\alpha _4\ast \mu_\alpha }{1+\frac{F^\alpha _4\ast \mu_\alpha }{k}})_{\alpha \in ]0,1[}$. There is a limit sequence $(g^l_1,g^l_2)$ in $(L^1([0,1]^2))^2$ such that up to subsequences $(f^{\alpha ,l}_1)$ (resp. $(f^{\alpha ,l}_2)$) converges to $g^l_1$ (resp. $g^l_2$) in $L^1([0,1]^2)$ when $\alpha \rightarrow 0$. They satisfy
\begin{align*}
&0\leq g^{2l}_1\leq g^{2l+2}_1\leq F^k_1\leq g^{2l+3}_1\leq g^{2l+1}_1,\\
&0\leq g^{2l}_2\leq g^{2l+2}_2\leq F^k_2\leq g^{2l+3}_2\leq g^{2l+1}_2,\quad l\in \N,\\
&\partial _xg^{2l+1}_1=  G-\frac{g^{2l+1}_1}{1+\frac{g^{2l+1}_1}{k}}\frac{g^{2l}_2}{1+\frac{g^{2l}_2}{k}},\quad \partial _xg^{2l}_1=  G-\frac{g^{2l}_1}{1+\frac{g^{2l}_1}{k}}\frac{g^{2l-1}_2}{1+\frac{g^{2l-1}_2}{k}},\\
&-\partial _xg^{2l+1}_2=  G-\frac{g^{2l}_1}{1+\frac{g^{2l}_1}{k}}\frac{g^{2l+1}_2}{1+\frac{g^{2l+1}_2}{k}},\quad -\partial _xg^{2l}_2=  G-\frac{g^{2l-1}_1}{1+\frac{g^{2l-1}_1}{k}}\frac{g^{2l}_2}{1+\frac{g^{2l}_2}{k}},\\
&g^l _1(0,y)= f_{b1}(y)\wedge \frac{k}{2},\quad g^l_2(1,y)= f_{b2}(y)\wedge \frac{k}{2},\quad y\in [0,1] ,
\end{align*}
where $G$ is the weak $L^1$ limit of $(\frac{F^\alpha _3}{1+\frac{F^\alpha _3}{k}}\frac{F^\alpha _4\ast \mu_\alpha }{1+\frac{F^\alpha _4\ast \mu_\alpha }{k}})_{\alpha \in ]0,1[}$ when $\alpha \rightarrow 0$. In particular, $(g^{2l}_1)_{l\in \N }$ and $(g^{2l}_2)_{l\in \N }$ (resp $(g^{2l+1}_1)_{l\in \N }$ and $(g^{2l+1}_2)_{l\in \N }$) non decreasingly (resp. non increasingly) converge in $L^1$ to some $g_1$ and $g_2$ (resp. $h_1$ and $h_2$) when $l\rightarrow +\infty $. The limits satisfy
\begin{align*}
&0\leq g_1\leq F^k_1\leq h_1,\quad 0\leq g_2\leq F^k_2\leq h_2,\\
&\partial _xh_1=  G-\frac{h_1}{1+\frac{h_1}{k}}\frac{g_2}{1+\frac{g_2}{k}},\quad \partial _xg_1=  G-\frac{g_1}{1+\frac{g_1}{k}}\frac{h_2}{1+\frac{h_2}{k}},\nonumber \\
&-\partial _xh_2=  G-\frac{g_1}{1+\frac{g_1}{k}}\frac{h_2}{1+\frac{h_2}{k}},\quad -\partial _xg_2=  G-\frac{h_1}{1+\frac{h_1}{k}}\frac{g_2}{1+\frac{g_2}{k}},\nonumber \\
&(h_1-g_1)(0,y)= 0,\quad (h_2-g_2)(1,y)= 0,\quad y\in [0,1] .
\end{align*}
Hence,
\begin{align*}
&(h_2-g_2)(x,y)= (h_1-g_1)(x,y)-(h_1-g_1)(1,y),\quad (x,y)\in [0,1] ^2,
\end{align*}
and
\begin{align*}
(h_1-g_1)(x,y)= &-(h_1-g_1)(1,y)\int _0^x\frac{h_1}{(1+\frac{h_1}{k})(1+\frac{g_2}{k})(1+\frac{h_2}{k})}(X,y)\\
&\exp \Big( -\int _X^x\frac{h_2(1+\frac{g_2}{k})-h_1(1+\frac{g_1}{k})}{(1+\frac{g_1}{k})(1+\frac{h_1}{k})(1+\frac{g_2}{k})(1+\frac{h_2}{k})}(r,y)dr\Big) dX.
\end{align*}
The non negativity of $h_1-g_1$, $g_1$, $g_2$, $h_1$ and $h_2$ implies that $h_1-g_1= 0$. The same holds for $h_2-g_2$. Consequently
\begin{align*}
&g_1= h_1= F^k_1,\quad g_2= h_2= F^k_2.
\end{align*}
$(F^\alpha _1)_{\alpha \in ]0,1[}$ converges to $F^k_1$ in $L^1([0,1]^2)$ when $\alpha \rightarrow 0$. Indeed, given $\eta >0$, choose $l_0$ big enough so that $\parallel g^{2l_0+1}_1-g^{2l_0}_1\parallel _{L^1}<\eta $ and $\parallel g^{2l_0}_1-F^k_1\parallel _{L^1}<\eta $, then $\alpha _0$ small enough so that
\begin{align*}
&\parallel f^{\alpha , 2l_0+1}_1-g^{2l_0+1}_1\parallel _{L^1}\leq \eta \quad \text{and}\quad \parallel f^{\alpha ,2l_0}_1-g^{2l_0}_1\parallel _{L^1}\leq \eta ,\quad \alpha \in ] 0,\alpha_0[ .
\end{align*}
Then split $\parallel F^\alpha _1-F^k_1\parallel _{L^1}$ as follows
\begin{align*}
\parallel F^\alpha _1-F^k_1\parallel _{L^1}&\leq \parallel F^\alpha _1-f^{\alpha ,2l_0}_1\parallel _{L^1}+\parallel f^{\alpha ,2l_0}_1-g^{2l_0}_1\parallel _{L^1}+\parallel g^{2l_0}_1-F^k_1\parallel _{L^1}\nonumber \\
& \leq \parallel f^{\alpha , 2l_0+1}_1-f^{\alpha ,2l_0}_1\parallel _{L^1}+2\eta \quad \text{by}\quad \eqref{order-f-alpha-l}\nonumber \\
& \leq \parallel f^{\alpha , 2l_0+1}_1-g^{2l_0+1}_1\parallel _{L^1}+\parallel g^{2l_0+1}_1-g^{2l_0}_1\parallel _{L^1}+\parallel g^{2l_0}_1-f_1^{\alpha ,2l_0}\parallel _{L^1}+2\eta \nonumber \\
&\leq 5\eta ,\quad \alpha \in ] 0,\alpha_0[ .
\end{align*}
The $L^1$ convergence of $(F^\alpha _i)_{k\in \N}$ to $F^k_i$, $2\leq i\leq 4$, can be proven similarly. Passing to the limit when $\alpha \rightarrow 0$ in \eqref{F1-alpha}-\eqref{bcF3-F4-alpha} is straightforward. And so, $F^k$ is a solution to \eqref{Fk-1}-\eqref{bcFk-3-Fk-4}. \cqfd\\
%
%
%
%
%
\section {Passage to the limit when $k\rightarrow +\infty $.}
\label{passage-to-the-limit}
 \setcounter{equation}{0}
 \setcounter{theorem}{0}
The study of the passage to the limit is split into six lemmas. In Lemma \ref{mass-entropy}, uniform bounds are obtained for mass, entropy and entropy production term of the approximations. Lemma \ref{lemma-df-Omega} splits $[0,1]^2$ into 
`large' sets of type $0\leq x\leq1$ times a 'large'  set in $y$ for $(F^k_1,F^k_2)$ (resp. a 'large'  set in $x$ times $0\leq y\leq 1$ for $(F^k_3,F^k_4)$), where the approximations are uniformly bounded in $L^\infty $, and their complements where the mass of the approximations is small. Lemma \ref{half-equi-integrabilite} proves uniform equicontinuity with respect to the $x$ (resp. $y$) variable of the two first (resp. last) components of the approximations. In Lemma \ref{strong-compactness-gain-term-of-F_1}, $L^1$-compactness of a truncated gain term of the approximations is proven. Lemma \ref{Fk-strong-compactness} proves that the approximations form  a Cauchy sequence in $L^1([0,1]^2)$. Their limit is proven to be a renormalized solution to the Broadwell model in Lemma \ref{F-renormalized-solution} .\\
In this section, $c_b$ denotes constant that only depend on the given boundary value $f_{b}$. 
 %
 %
\begin{lemma}\label{mass-entropy}
\hspace*{1.in}\\
There are constants $c_b$  such that
\begin{align}
&\int F^k_i(x,y)dxdy\leq {c_b},\label{mass}\\
&\int _{F^k_i(x,y)>k}F^k_i(x,y)dxdy\leq \frac{c_b}{\ln k},\quad i\in \{ 1,\cdot \cdot \cdot ,4\} ,\label{entropy}\\
&\int (\frac{F^k_1}{1+\frac{F^k_1}{k}}\frac{F^k_2}{1+\frac{F^k_2}{k}}-\frac{F^k_3}{1+\frac{F^k_3}{k}}\frac{F^k_4}{1+\frac{F^k_4}{k}})\ln \frac{F^k_1F^k_2(1+\frac{F^k_3}{k})(1+\frac{F^k_4}{k})}{(1+\frac{F^k_1}{k})(1+\frac{F^k_2}{k})F^k_3F^k_4}(x,y)dxdy\leq {c_b},\quad k>2.\label{production-entropy}
\end{align}
\end{lemma}
\underline{Proof of Lemma \ref{mass-entropy}.}\\
Adding \eqref{Fk-1}-\eqref{Fk-4}, integrating the resulting equation on $[0,1]^2$ and taking \eqref{bcFk-1-Fk-2}-\eqref{bcFk-3-Fk-4} into account, implies that total outflow equals total inflow. Also using $\partial_x(F_1^k+F_2^k)=\partial_ y(F_3^k+F_4^k)=0$ implies boundedness of the total mass $\sum _{i=1}^4\int F^k_i(x,y)dxdy$.  Multiply  \eqref{Fk-1} (resp.  \eqref{Fk-2}, resp.  \eqref{Fk-3}, resp.  \eqref{Fk-4}) by $\ln \frac{F^k_1}{1+\frac{F^k_1}{k}}$ (resp. $\ln \frac{F^k_2}{1+\frac{F^k_2}{k}}$, resp. $\ln \frac{F^k_3}{1+\frac{F^k_3}{k}}$, resp. $\ln \frac{F^k_4}{1+\frac{F^k_1}{4}}$), add the corresponding equations, and integrate the resulting equation on $[0,1]^2$. Denoting by $D^k$ the entropy production term for the approximation $F^k$,
\begin{align*}
&D^k= \int (\frac{F^k_1}{1+\frac{F^k_1}{k}}\frac{F^k_2}{1+\frac{F^k_2}{k}}-\frac{F^k_3}{1+\frac{F^k_3}{k}}\frac{F^k_4}{1+\frac{F^k_4}{k}})\ln \frac{F^k_1F^k_2(1+\frac{F^k_3}{k})(1+\frac{F^k_4}{k})}{(1+\frac{F^k_1}{k})(1+\frac{F^k_2}{k})F^k_3F^k_4}(x,y)dxdy,
\end{align*}
leads to
\begin{align*}
&\int \Big( F^k_1\ln F^k_1-k(1+\frac{F^k_1}{k})\ln (1+\frac{F^k_1}{k})\Big) (1,y)dy+ \int \Big( F^k_2\ln F^k_2-k(1+\frac{F^k_2}{k})\ln (1+\frac{F^k_2}{k})\Big) (0,y)dy\\
&+\int \Big( F^k_3\ln F^k_3-k(1+\frac{F^k_3}{k})\ln (1+\frac{F^k_3}{k})\Big) (x,1)dx+ \int \Big( F^k_4\ln F^k_4-k(1+\frac{F^k_4}{k})\ln (1+\frac{F^k_4}{k})\Big) (x,0)dx\\
&+D^k\leq c_b.
\end{align*}
Moreover, 
\begin{align*}
k\int \ln (1+\frac{F^k_i}{k})&\leq \int F^k_i\leq c_b,\quad 1\leq i\leq 4.
\end{align*}
Hence
\begin{align*}
&\int \Big( F^k_1\ln \frac{F^k_1}{1+\frac{F^k_1}{k}}(1,y)+F^k_2\ln \frac{F^k_2}{1+\frac{F^k_2}{k}}(0,y)\Big) dy+\int \Big( F^k_3\ln \frac{F^k_3}{1+\frac{F^k_3}{k}}(x,1)+F^k_4\ln \frac{F^k_4}{1+\frac{F^k42}{k}}(x,0)\Big) dx\\
&+D^k\leq c_b.
\end{align*}
Consequently,
\begin{align*}
&\int _{F^k_1(1,y)>\frac{k}{k-1}}F^k_1\ln \frac{F^k_1}{1+\frac{F^k_1}{k}}(1,y)dy+ \int _{F^k_2(0,y)>\frac{k}{k-1}}F^k_2\ln \frac{F^k_2}{1+\frac{F^k_2}{k}}(0,y)dy\\
&+\int _{F^k_3(x,1)>\frac{k}{k-1}}F^k_3\ln \frac{F^k_3}{1+\frac{F^k_3}{k}}(x,1)dx+ \int _{F^k_4(x,0)>\frac{k}{k-1}}F^k_4\ln \frac{F^k_4}{1+\frac{F^k_4}{k}}(x,0)dx+D_k\\
&\leq c_b,\quad k> 2.
\end{align*}
And so, \eqref{production-entropy} holds. Moreover, for any $\Lambda >2$ and $k>2$,
\begin{align}\label{entropy-flux}
&\ln \frac{\Lambda }{1+\frac{\Lambda }{k}}\Big( \int _{F^k_1(1,y)>k}F^k_1(1,y)dy+\int _{F^k_2(0,y)>k}F^k_2(0,y)dy\nonumber \\
&\hspace*{0.6in}+\int _{F^k_3(x,1)>k}F^k_3(x,1)dx+\int _{F^k_4(x,0)>k}F^k_4(x,0)dx\Big) \nonumber \\
&\leq c_b+\int _{F^k_1(1,y)<\frac{k}{k-1}}F^k_1\mid \ln \frac{F^k_1}{1+\frac{F^k_1}{k}}\mid (1,y)dy+\int _{F^k_2(0,y)<\frac{k}{k-1}}F^k_2\mid \ln \frac{F^k_2}{1+\frac{F^k_2}{k}}\mid (0,y)dy\nonumber \\
&\hspace*{0.6in}+\int _{F^k_3(x,1)<\frac{k}{k-1}}F^k_3\mid \ln \frac{F^k_3}{1+\frac{F^k_3}{k}}\mid (x,1)dx+\int _{F^k_4(x,0)<\frac{k}{k-1}}F^k_4\mid \ln \frac{F^k_4}{1+\frac{F^k_4}{k}}\mid (x,0)dx\nonumber \\
&\leq c_b+2,\quad k>2.
\end{align}
In particular,
\begin{align}\label{entropy-ter}
&\int _{F^k_1(1,y)>k}F^k_1(1,y)dy+\int _{F^k_2(0,y)>k}F^k_2(0,y)dy\nonumber \\
&+\int _{F^k_3(x,1)>k}F^k_3(x,1)dx+\int _{F^k_4(x,0)>k}F^k_4(x,0)dx\leq \frac{c_b}{\ln k},\quad k>2.
\end{align}
Since 
\begin{align}\label{sumF1F2}
&(F^k_1+F^k_2)(x,y)= F^k_1(1,y)+f_{b2}(y)\wedge \frac{k}{2},\quad (x,y)\in [0,1]^2,
\end{align}
it holds that
\begin{align*}
&F^k_1(x,y)>k\Rightarrow F^k(1,y)>\frac{k}{2},\quad (x,y)\in [0,1]^2.
\end{align*}
Consequently, for some subset $\omega _k$ of $[0,1]$ such that $\lvert \omega _k\rvert <\frac{c}{k}$,
\begin{align*}
\int _{F^k_1(x,y)>k}F^k_1(x,y)dxdy&\leq \int _{F^k_1(1,y)>\frac{k}{2}}F^k_1(1,y)dy+\int _{\omega _k}f_{b2}(y)dy\nonumber \\
&\leq \frac{c}{\ln k},
\end{align*}
by \eqref{entropy-flux} and {the boundedness of the $f_{b2}$ entropy.} \cqfd
\hspace*{1.in}\\
\hspace*{1.in}\\
%
%
%
%
\begin{lemma}\label{lemma-df-Omega}
\hspace*{1.in}\\
For $\epsilon >0$, $\Lambda \geq \exp (\frac{2c_b}{\epsilon })$ and $k\geq \exp (\frac{3c_b}{\epsilon })$, there is a subset $\Omega ^{\epsilon  \Lambda }_{k1}$ of $[0,1] $ with measure smaller than $\frac{c_b\epsilon }{\Lambda }$ such that
\begin{align}
&F^k_{1}(x,y)\leq \frac{\Lambda }{\epsilon }\exp (\frac{2\Lambda }{\epsilon }),\quad F^k_2(x,y)\leq \frac{2\Lambda }{\epsilon }\exp (\frac{2\Lambda }{\epsilon }),\quad x\in [0,1] ,\quad y\in [0,1] \setminus \Omega ^{\epsilon \Lambda }_{k1},\label{bdd-char-1}\\ 
&\int _0^1\Big( \int _{\Omega ^{\epsilon \Lambda }_{k1}}(F^k_1+F^k_2)(x,y)dy\Big) dx\leq c_b\epsilon .\label{small-mass}
\end{align}
\end{lemma}
\underline{Proof of Lemma \ref{lemma-df-Omega}.}\\
Since $f_{b2}\in L^1([0,1])$ and
\begin{align*}
&\int _0^1(F^k_1(1,y)+F^k_2(0,y))dy+\int _0^1(F^k_3(x,1)+F^k_4(x,0))dx\leq c_b,
\end{align*}
the measure of the set 
\begin{align}\label{df-Omega}
&\Omega ^{\epsilon \Lambda }_{k1}:= \{ y\in [0,1] ; f_{b2}(y)\geq \frac{\Lambda }{\epsilon } \text{   or   } F^k_1(1,y)\geq \frac{\Lambda }{\epsilon } \} ,
\end{align}
is smaller than $\frac{c_b\epsilon }{\Lambda }$. $(F^k_1, F^k_2)$ is uniformly bounded on $[0,1] \times ([0,1] \setminus \Omega ^{\epsilon \Lambda }_{k1})$, since 
\begin{align*}
F^k_1(x,y)&\leq F^k_{1}(1,y)\exp (\int _0^1F^k_{2}(X,y)dX)\\
&\leq F^k_{1}(1,y)\exp (F^k_1(1,y)+f_{b2}(y))\quad \text{by}\quad \eqref{sumF1F2}\\
&\leq \frac{\Lambda }{\epsilon }\exp (\frac{2\Lambda }{\epsilon }),
\end{align*}
and
\begin{align*}
F^k_2(x,y)&\leq F^k_{2}(0,y)\exp (\int _0^1F^k_{1}(X,y)dX)\\
&\leq  (F^k_{1}(1,y)+f_{b2}(y))\exp (F^k_1(1,y)+f_{b2}(y))\\
&\leq \frac{2\Lambda }{\epsilon }\exp (\frac{2\Lambda }{\epsilon }),\quad x\in [0,1], \quad y\in [0,1] \setminus \Omega ^{\epsilon \Lambda }_{k1}.
\end{align*}
Moreover, for any $\Lambda \geq \exp (\frac{2c_b}{\epsilon })$ and $k\geq \exp (\frac{3c_b}{\epsilon })$,
\begin{align*}
&\int _0^1\Big( \int _{\Omega ^{\epsilon \Lambda }_{k1}}(F^k_1+F^k_2)(x,y)dy\Big) dx= \int _{\Omega ^{\epsilon \Lambda }_{k1}}(F^k_1(1,y)+f_{b2}(y))dy\\
&\leq  \int _{y\in \Omega ^{\epsilon \Lambda }_{k1}; F^k_1(1,y)<\Lambda }F^k_1(1,y)dy+\int _{F^k_1(1,y)>\Lambda }F^k_1(1,y)dy\\
&+\int _{y\in \Omega ^{\epsilon \Lambda }_{k1}, f_{b2}(y)<\Lambda }f_{b2}(y)dy+\int _{f_{b2}(y)>\Lambda }f_{b2}(y)dy\\
&\leq 2\Lambda \lvert \Omega ^{\epsilon \Lambda }_{k1}\rvert +\frac{c_b}{\ln \frac{\Lambda }{1+\frac{\Lambda }{k}}}+\frac{c_b}{\ln \Lambda }\quad \text{by   }\eqref{entropy-flux}\hspace*{0.03in} \text{and   }\eqref{hyp-fb}\\
&\leq c_b\epsilon .
\end{align*}
\cqfd
%
 %
 %
 %
\begin{lemma}\label{half-equi-integrabilite}
\hspace*{1.in}\\
There is $c_b>0$, and for $\epsilon >0$ given there is $\delta >0$ such that for $\lvert h\rvert < \delta $, uniformly in $k\in \N ^*$,
\begin{align}\label{equi-integrated-in-time-of-functions}
&\int _{[0,1]^2}\lvert F^k_i(x+h,y)-F^k_i(x,y)\rvert dxdy\leq c_b\epsilon ,\quad i\in \{ 1,2\} ,\nonumber \\
&\int _{[0,1]^2}\lvert F^k_i(x,y+h)-F^k_i(x,y)\rvert dxdy\leq c_b\epsilon ,\quad i\in \{ 3,4\} .
\end{align}
\end{lemma}
\underline{Proof of Lemma \ref{half-equi-integrabilite}.}\\
The four cases $F^k_1$,..., $F^k_4$ are analogous. The detailed estimates are carried out for $F^k_1$.
The translational  $L^1$ equicontinuity in the $x$-direction for $\ln (1+F^k_1)$ is obtained as follows from the $\partial_x$-term in the renormalized equation. Consider $h\in [0,1[ $. Write the equation for $F^k_1$ in renormalized form \eqref{renormalized-form} integrated on $[x,x+h]$, where the integration from  $x+h>1$  tending to zero with $h$ uniformly in $k$, is being omitted from the following computations;
 \begin{align}
&\ln (1+F^k_1(x+h,y))-\ln (1+F^k_1(x,y))\nonumber \\
&= \int_x^{x+h} \frac{1}{1+F^k_1}\Big( \frac{F^k_3}{1+\frac{F^k_3}{k}}\frac{F^k_4}{1+\frac{F^k_4}{k}}-\frac{F^k_1}{1+\frac{F^k_1}{k}}\frac{F^k_2}{1+\frac{F^k_2}{k}}\Big) (X,y)dX.
\end{align}
Denote by $\rm{sgn}$ the sign function, $\rm{sgn}(r)= 1$ if $r>0$, $\rm{sgn}(r)= -1$ if $r<0$. Multiply the previous equation by $\rm{sgn}\big( \ln (1+F^k_1(x+h,y))-\ln (1+F^k_1(x,y))\big) $ and integrate on $[0,1]^2$. Uniformly w.r.t. $k\in \N ^*$,
\begin{align}\label{pf-lemma-22-a}
&\int \lvert \ln (F^k_1(x+h,y)+1)-\ln (F^k_1(x,y)+1)\rvert dxdy\nonumber \\
&\leq h\int_{[0,1]^2}\frac{\lvert \frac{F^k_3}{1+\frac{F^k_3}{k}}\frac{F^k_4}{1+\frac{F^k_4}{k}}-\frac{F^k_1}{1+\frac{F^k_1}{k}}\frac{F^k_2}{1+\frac{F^k_2}{k}}\rvert }{(1+F^k_1)}(X,y)dXdy\nonumber \\
&\leq h\Big( \int_{\frac{F^k_3}{1+\frac{F^k_3}{k}}\frac{F^k_4}{1+\frac{F^k_4}{k}}<\frac{F^k_1}{1+\frac{F^k_1}{k}}\frac{F^k_2}{1+\frac{F^k_2}{k}}}\frac{F^k_1}{(1+F^k_1)(1+\frac{F^k_1}{k})}\frac{F^k_2}{1+\frac{F^k_2}{k}}(X,y)dXdy\nonumber \\
&\hspace*{0.22in}+\int_{\frac{F^k_3}{1+\frac{F^k_3}{k}}\frac{F^k_4}{1+\frac{F^k_4}{k}}>\frac{F^k_1}{1+\frac{F^k_1}{k}}\frac{F^k_2}{1+\frac{F^k_2}{k}}}\frac{F^k_3}{(1+F^k_1)(1+\frac{F^k_3}{k})}\frac{F^k_4}{1+\frac{F^k_4}{k}}(X,y)dXdy\Big) \nonumber \\
&\leq h\Big( \int F^k_2(X,y)dXdy+\int \frac{F^k_3}{(1+F^k_1)(1+\frac{F^k_3}{k})}\frac{F^k_4}{1+\frac{F^k_4}{k}}(X,y)dXdy\Big) \nonumber \\
&\leq h\Big( c_b+\int_{\frac{F^k_3}{1+\frac{F^k_3}{k}}\frac{F^k_4}{1+\frac{F^k_4}{k}}<2\frac{F^k_1}{1+\frac{F^k_1}{k}}\frac{F^k_2}{1+\frac{F^k_2}{k}}}\frac{F^k_3}{(1+F^k_1)(1+\frac{F^k_3}{k})}\frac{F^k_4}{1+\frac{F^k_4}{k}}(X,y)dXdy\nonumber\\
&\hspace*{0.5in}+\int_{\frac{F^k_3}{1+\frac{F^k_3}{k}}\frac{F^k_4}{1+\frac{F^k_4}{k}}>2\frac{F^k_1}{1+\frac{F^k_1}{k}}\frac{F^k_2}{1+\frac{F^k_2}{k}}}\frac{F^k_3}{1+\frac{F^k_3}{k}}\frac{F^k_4}{1+\frac{F^k_4}{k}}(X,y)dXdy\Big) \nonumber\\
&\leq c_bh(1+(\ln 2)^{-1})\leq c_bh.
\end{align}
Recall that for any non negative real numbers $x_1>x_2$, there is $\theta \in ]0,1[$ such that
\begin{align*}
x_1-x_2&= \exp (\ln (1+x_1))-\exp (\ln (1+x_2))\\
&=\exp \big( \theta \ln(1+x_1)+(1-\theta )\ln(1+x_2)\big) \big( \ln (1+x_1)-\ln (1+x_2)\big) .
\end{align*}
 And so  the $L^1$-norms of the translation differences of $F^k_1$ and $\ln (1+F^k_1)$, are equivalent on $[0,1] \times ([0,1] \setminus \Omega ^{\epsilon \Lambda }_{k1})$
since $F^k_1$ and $(x,y)\rightarrow F^k_1(x+h,y)$ are bounded in $L^\infty ([0,1] \times ([0,1] \setminus \Omega ^{\epsilon \Lambda }_{k1}))$. 
%
There is also the small set with mass bounded by $\epsilon$, where $(x,y)\rightarrow F^k_1(x+h,y)$ is not in $\Omega ^{\epsilon \Lambda }_{k1}$.
Together with \eqref{pf-lemma-22-a} this proves the translational equicontinuity in the $x$-direction for $k\geq \exp (\frac{3c_b}{\epsilon })$. The proof for $h\in ]-1,0[ $ is similar. \cqfd
%
%
%
%
Given $\epsilon >0$, $\Lambda \geq \exp (\frac{2c_b}{\epsilon })$ and $k \geq \exp (\frac{3c_b}{\epsilon })$, let $\Omega ^{\epsilon \Lambda }_{k1}\subset [0,1]$ as defined in Lemma \ref{lemma-df-Omega}, and take $\chi^{\epsilon \Lambda }_{k1}$ as the corresponding cutoff function, 
\begin{align*}
&\chi^{\epsilon \Lambda }_{k1}(y)= 1\text{   if}\hspace*{0.06in}y\notin \Omega _{k1}^{\epsilon \Lambda },\quad \quad \chi^{\epsilon \Lambda }_{k1}(y)= 0\text{   if}\hspace*{0.06in}y\in \Omega_{k1}^{\epsilon \Lambda}.
\end{align*}
%
%
\begin{lemma}\label{strong-compactness-gain-term-of-F_1}
\hspace*{1.in}\\
Let $(\alpha ^k)_{k\in \N }$ be a non negative sequence bounded in $L^\infty $ and compact in $L^1$  . The sequences
\begin{align*}
&\Big( \chi_{k1}^{\epsilon \Lambda}(y)\int _0^x\frac{F^k_3}{1+\frac{F^k_3}{k}}\frac{F^k_4}{1+\frac{F^k_4}{k}}(X,y)e^{-\int _X^x\alpha ^k(u,y)du}dX\Big) _{k\in \N ^*}\\
&\text{and    }\Big( \chi_{k1}^{\epsilon \Lambda}(y)\int _x^1\frac{F^k_3}{1+\frac{F^k_3}{k}}\frac{F^k_4}{1+\frac{F^k_4}{k}}(X,y)e^{-\int _x^X\alpha ^k(u,y)du}dX\Big) _{k\in \N ^*},\\
&\quad \Big( \text{resp.   }\big( \chi_{k1}^{\epsilon \Lambda}(y)\int _0^1\frac{F^k_3}{1+\frac{F^k_3}{k}}\frac{F^k_4}{1+\frac{F^k_4}{k}}(X,y)dX\big) _{k\in \N ^*}\Big) ,
\end{align*}
are compact in $L^1([0,1]^2)$ (resp. in $L^1([0,1]))$.
\end{lemma}
\underline{Proof of Lemma \ref{strong-compactness-gain-term-of-F_1}.}\hspace*{0.05in}For any $\gamma >1$, using Lemmas \ref{mass-entropy}-\ref{lemma-df-Omega},
\begin{align*}
& \int \chi_{k1}^{\epsilon \Lambda}(y)\lvert \int _0^{x+h}\frac{F^k_3}{1+\frac{F^k_3}{k}}\frac{F^k_4}{1+\frac{F^k_4}{k}}(X,y)e^{-\int _X^{x+h}\alpha ^k(u,y)du}dX\\
&\hspace*{0.7in}-\int _0^x\frac{F^k_3}{1+\frac{F^k_3}{k}}\frac{F^k_4}{1+\frac{F^k_4}{k}}(X,y)e^{-\int _X^x\alpha ^k(u,y)du}dX\rvert dxdy\\
&\leq \int \chi_{k1}^{\epsilon \Lambda}(y)\lvert \int _x^{x+h}\frac{F^k_3}{1+\frac{F^k_3}{k}}\frac{F^k_4}{1+\frac{F^k_4}{k}}(X,y)dX\rvert dxdy\\
&\hspace*{0.7in}+\int \chi_{k1}^{\epsilon \Lambda}(y)\int _0^x\frac{F^k_3}{1+\frac{F^k_3}{k}}\frac{F^k_4}{1+\frac{F^k_4}{k}}(X,y)dX\lvert \int _x^{x+h}\alpha ^k(u,y)du\rvert dxdy\\
&\leq \frac{c_b}{\ln \gamma }+\gamma h\int \chi_{k1}^{\epsilon \Lambda}(y)F^k_1F^k_2(x,y)dxdy\\
&\leq \frac{c_b}{\ln \gamma }+2\gamma h\big( \frac{\Lambda }{\epsilon }\big) ^2e^{\frac{4\Lambda }{\epsilon }}.
\end{align*} 
Choosing $\gamma $ big enough, then $h$ small enough, proves the translational $L^1$ equicontinuity in the $x$ direction of $\Big( \chi_{k1}^{\epsilon \Lambda}(y)\int _0^x\frac{F^k_3}{1+\frac{F^k_3}{k}}\frac{F^k_4}{1+\frac{F^k_4}{k}}(X,y)e^{-\int _X^x\alpha ^k(u,y)du}dX\Big) _{k\in \N ^*}$. Let us prove its translational $L^1$ equicontinuity in the $y$ direction. Given $\tilde{\epsilon }>0$, let 
\begin{align}
&\gamma >\exp (\frac{3c_b}{\tilde{\epsilon }}),\quad \epsilon _3<\frac{\tilde{\epsilon }}{6c_b\gamma }\big( \frac{\epsilon }{\Lambda }\big) ^2e^{-\frac{4\Lambda }{\epsilon }},\quad \Lambda _3\geq \exp (\frac{2c_b}{\epsilon _3}).
\end{align}
Let $\Omega ^{\epsilon _3\Lambda _3}_{k3}\subset [0,1]$ as defined in Lemma \ref{lemma-df-Omega} for $(F^k_3,F^k_4)$, and $\chi^{\epsilon _3\Lambda _3}_{k3}$ the corresponding cutoff function, 
\begin{align*}
&\chi^{\epsilon _3\Lambda _3}_{k3}(x)= 1\text{   if}\hspace*{0.08in}x\notin  \Omega _{k3}^{\epsilon _3\Lambda _3},\quad \quad \chi^{\epsilon \Lambda}_{k3}(x)= 0\text{   if}\hspace*{0.08in}x\in \Omega_{k3}^{\epsilon _3\Lambda _3}.
\end{align*}
First,
\begin{align*}
&\int \Big( \int _{X\in [0,x]; \frac{F^k_3}{1+\frac{F^k_3}{k}}\frac{F^k_4}{1+\frac{F^k_4}{k}}(X,y)>\gamma \frac{F^k_1}{1+\frac{F^k_1}{k}}\frac{F^k_2}{1+\frac{F^k_2}{k}}(X,y)}\frac{F^k_3}{1+\frac{F^k_3}{k}}\frac{F^k_4}{1+\frac{F^k_4}{k}}(X,y)dX\Big) dxdy\leq \frac{c_b}{\ln \gamma }\leq \frac{\tilde{\epsilon }}{3}.
\end{align*}
Moreover,
\begin{align*}
&\int \chi ^{\epsilon \Lambda }_{k1}(y)\int _{X\in [0,x]; \frac{F^k_3}{1+\frac{F^k_3}{k}}\frac{F^k_4}{1+\frac{F^k_4}{k}}(X,y)<\gamma \frac{F^k_1}{1+\frac{F^k_1}{k}}\frac{F^k_2}{1+\frac{F^k_2}{k}}(X,y)}(1-\chi^{\epsilon _3\Lambda _3}_{k3}(X))\frac{F^k_3}{1+\frac{F^k_3}{k}}\frac{F^k_4}{1+\frac{F^k_4}{k}}(X,y)dXdxdy\\
&\leq 2c_b\gamma \Big( \frac{\Lambda }{\epsilon }\Big) ^2e^{\frac{4\Lambda }{\epsilon }}\epsilon _3\\
&\leq \frac{\tilde{\epsilon }}{3},\quad \text{by the definition of    }\epsilon _3.
\end{align*}
Given the boundedness of $(F^k_3,F^k_4)_{k\geq \exp (\frac{3c_b}{\epsilon _3})}$ on $\big( \Omega _{k3}^{\epsilon _3\Lambda _3}\big) ^c\times [0,1]$, and the statements of Lemmas \ref{lemma-df-Omega}-\ref{half-equi-integrabilite} for $(F^k_3,F^k_4)$, there is $h_3>0$ such that
\begin{align*}
&\int \int _0^x\chi^{\epsilon _3\Lambda _3}_{k3}(X)\lvert \frac{F^k_3}{1+\frac{F^k_3}{k}}\frac{F^k_4}{1+\frac{F^k_4}{k}}(X,y+h)-\frac{F^k_3}{1+\frac{F^k_3}{k}}\frac{F^k_4}{1+\frac{F^k_4}{k}}(X,y)\rvert dXdxdy\leq \frac{\tilde{\epsilon }}{3},
\end{align*}
for $h\in ]0,h_3[$, uniformly with respect to $k\geq \exp (\frac{3c_b}{\epsilon _3})$.\\
The proofs of the $L^1([0,1]^2)$ (resp. $L^1([0,1])$) compactness of 
\begin{align*}
&\Big( \chi_{k1}^{\epsilon \Lambda}(y)\int _x^1\frac{F^k_3}{1+\frac{F^k_3}{k}}\frac{F^k_4}{1+\frac{F^k_4}{k}}(X,y)e^{-\int _x^X\alpha ^k(u,y)du}dX\Big) _{k\in \N ^*},\\
&\Big(\text{resp.} \big( \chi_{k1}^{\epsilon \Lambda}(y)\int _0^1\frac{F^k_3}{1+\frac{F^k_3}{k}}\frac{F^k_4}{1+\frac{F^k_4}{k}}(X,y)dX\big) _{k\in \N ^*}\Big) 
\end{align*}
are similar. \cqfd
%
%
%
\begin{lemma}\label{Fk-strong-compactness}
\hspace*{1.in}\\
$(F^k)_{k\in \N ^*}$ is compact in $L^1([0,1]^2)$. \\
\end{lemma}
\underline{Proof of Lemma \ref{Fk-strong-compactness}.}\\
By \eqref{mass}-\eqref{entropy}, $(F^k)_{k\in \N ^*}$ is weakly compact in $(L^1([0,1]^2))^4$. Denote by $F$ the weak limit of a subsequence, still denote $(F^k)$. Let us prove that $(F^k_1)_{k\in \N ^*}$ is strongly compact in $L^1([0,1]^2)$. It is by \eqref{small-mass} enough to prove that up to a subsequence, given $\epsilon >0$, for $\Lambda \geq e^{\frac{2c_b}{\epsilon }}$, $k\geq e^{\frac{3c_b}{\epsilon }}$ and $\Omega ^{\epsilon \Lambda }_{k1}$ as defined in Lemma 3.2,
$(\chi ^{\epsilon \Lambda }_{k1}F^k_1)_{k\in \N ^*}$ is strongly compact in $L^1([0,1]^2)$. 
For every $F^k $ in the subsequence, consider the approximation scheme $(f^{k,l}_1, f^{k,l}_2)_{l\in \N }$ of $(F^k_1, F^k_2)$, defined by
\begin{align}
&f^{k,-1}_1=f^{k,-1}_2=f^{k,0}_1=f^{k,0}_2= 0,\nonumber \\
&f^{k,l+1}_1(x,y)= f_{b1}(y)+\int _0^x\big( \chi ^{\epsilon \Lambda }_{k1}(y) \frac{F^k_3}{1+\frac{F^k_3}{k}}\frac{F^k_4}{1+\frac{F^k_4}{k}}-\frac{f^{k,l+1}_1}{1+\frac{f^{k,l-1}_1}{k}}\frac{f^{k,l}_2}{1+\frac{f^{k,l}_2}{k}}\big) (X,y)dX,\label{approx-scheme-delta-1a}\\
&f^{k,l+1}_2(x,y)= f_{b2}(y)+\int _x^1\big( \chi ^{\epsilon \Lambda }_{k1}(y) \frac{F^k_3}{1+\frac{F^k_3}{k}}\frac{F^k_4}{1+\frac{F^k_4}{k}}-\frac{f^{k,l}_1}{1+\frac{f^{k,l}_1}{k}}\frac{f^{k,l+1}_2}{1+\frac{f^{k,l-1}_2}{k}}\big) (X,y)dX.\label{approx-scheme-delta-1b}
\end{align}
By induction on $l$, and using an exponential form of $(f_1^{k,l+1},f_2^{k,l+1})$, it holds that
\begin{align}\label{approx-scheme-delta-2}
&f^{k,2l}_1\leq f^{k,2l+2}_1,\quad f^{k,2l+3}_1\leq f^{k,2l+1}_1,\nonumber \\
&f^{k,2l}_2\leq f^{k,2l+2}_2,\quad f^{k,2l+3}_2\leq f^{k,2l+1}_2,\quad (x,y)\in [0,1]^2,\quad k\in \N ^*,\quad l\in \N ,
\end{align} 
and
\begin{align}\label{approx-scheme-delta-2bis}
&f^{k,2l}_1\leq F^k_1\leq f^{k,2l+1}_1,\quad f^{k,2l}_2\leq F^k_2\leq f^{k,2l+1}_2,\quad (x,y)\in [0,1] \times (\Omega ^{\epsilon \Lambda }_{k1})^c,\quad k\in \N ^*,\quad l\in \N .
\end{align} 
The sequence $(\chi ^{\epsilon \Lambda }_{k1}f^{k,2l}_1)_{k\geq e^{\frac{3c_b}{\epsilon }}}$ (resp. $(\chi ^{\epsilon \Lambda }_{k1}f^{k,2l}_2)_{k\geq e^{\frac{3c_b}{\epsilon }}}$) is bounded from above by 
$(\chi ^{\epsilon \Lambda }_{k1}F^k_1)_{k\geq e^{\frac{3c_b}{\epsilon }}}$ (resp. $(\chi ^{\epsilon \Lambda }_{k1}F^k_2)_{k\geq e^{\frac{3c_b}{\epsilon }}}$), hence by $\frac{2\Lambda }{\epsilon }\exp (\frac{2\Lambda }{\epsilon })$. The sequence $(\chi ^{\epsilon \Lambda }_{k1}f^{k,2l+1}_1)_{k\geq e^{\frac{3c_b}{\epsilon }}}$ (resp. $(\chi ^{\epsilon \Lambda }_{k1}f^{k,2l+1}_2)_{k\geq e^{\frac{3c_b}{\epsilon }}}$) is bounded by $\frac{2\Lambda }{\epsilon }\exp (\frac{2\Lambda }{\epsilon })(1+\frac{2\Lambda }{\epsilon }\exp (\frac{2\Lambda }{\epsilon }))$, since
\begin{align*}
&\chi ^{\epsilon \Lambda }_{k1}(y)f^{k,2l+1}_1(x,y)= \chi ^{\epsilon \Lambda }_{k1}(y)F^k_1(x,y)\\
&+\chi ^{\epsilon \Lambda }_{k1}(y)\int _0^x\frac{F^k_1}{1+\frac{F^k_1}{k}}(\frac{F^k_2}{1+\frac{F^k_2}{k}}-\frac{f^{k,l}_2}{1+\frac{f^{k,l}_2}{k}})(X,y)e^{-\int _X^x\frac{f^{k,l}_2}{(1+\frac{f^{k,l}_2}{k})(1+\frac{f^{k,l-1}_1}{k})(1+\frac{F^{k}_1}{k})}(r,y)dr}dX\\
&\hspace*{1.2in}\leq \chi ^{\epsilon \Lambda }_{k1}F^k_1(x,y)+\chi ^{\epsilon \Lambda }_{k1}(y)\int _0^xF^k_1F^k_2(X,y)dX.
\end{align*}
By Lemma \ref{strong-compactness-gain-term-of-F_1}, there is a subsequence of $\big( \chi_{k1}^{\epsilon \Lambda}(y)\int _0^1\frac{F^k_3}{1+\frac{F^k_3}{k}}\frac{F^k_4}{1+\frac{F^k_4}{k}}(X,y)dX\big) _{k\in \N ^*}$, still denoted by \\
$\big( \chi_{k1}^{\epsilon \Lambda}(y)\int _0^1\frac{F^k_3}{1+\frac{F^k_3}{k}}\frac{F^k_4}{1+\frac{F^k_4}{k}}(X,y)dX\big) _{k\in \N ^*}$, converging in $L^1([0,1])$ to some $\tilde{F}_1$. Given $\eta >0$, there is a subset $\omega _\eta $ of $[0,1]$ with measure smaller than $\eta $ such that on $\omega _\eta ^c$ the convergence of this sequence is uniform and $(\tilde{F}_1,f_{b1}, f_{b2})$ is bounded. It follows from \eqref{approx-scheme-delta-1a}-\eqref{approx-scheme-delta-1b} and the non-negativity of $(f_1^{k,2l},f_2^{k,2l})_{(k,l)\in \N ^2}$ that $(f_1^{k,2l},f_2^{k,2l})_{(k,l)\in \N ^2}$ is bounded on $[0,1] \times \omega _\eta ^c$. Given these bounds, Lemma \ref{strong-compactness-gain-term-of-F_1} and the expression of $(f_1^{k,l},f_2^{k,l})$ in exponential form, it holds by induction that for each $l\in \N $, the sequence $(f^{k,l}_1,f^{k,l}_2)_{k\geq e^{\frac{3c_b}{\epsilon }}}$ is strongly compact in $L^1([0,1] \times \omega _\eta ^c)$. Denote by $(g^l_1, g^l_2)$ its limit up to a subsequence. By Lemma \ref{strong-compactness-gain-term-of-F_1}, let $G$ (resp. $H$)
with $\partial _xG= -\partial _xH$, be the limit in $L^1$ when $k\rightarrow +\infty $ of
\begin{align*}
&(\chi ^{\epsilon \Lambda }_{k1}(y)\int _0^x\frac{F^k_3}{1+\frac{F^k_3}{k}}\frac{F^k_4}{1+\frac{F^k_4}{k}}(X,y)dX)_{k\geq e^{\frac{3c_b}{\epsilon }}},\hspace*{0.01in} \big( \text{resp.    }(\chi ^{\epsilon \Lambda }_{k1}(y)\int _x^1\frac{F^k_3}{1+\frac{F^k_3}{k}}\frac{F^k_4}{1+\frac{F^k_4}{k}}(X,y)dX)_{k\geq e^{\frac{3c_b}{\epsilon }}}\big) .
\end{align*} 
$(g^{2l}_1, g^{2l}_2, g^{2l+1}_1, g^{2l+1}_2)$ satisfies
\begin{align}
& g^0_1= g^0_2= 0,\nonumber \\
&g^{2l}_1(x,y)=  f_{b1}(y)+G(x,y)-\int _0^xg^{2l}_1g^{2l-1}_2(X,y)dX,\quad l\in \N ^*,\nonumber \\
&g^{2l+1}_1(x,y)=  f_{b1}(y)+G(x,y)-\int _0^xg^{2l+1}_1g^{2l}_2(X,y)dX,\quad l\in \N ,\nonumber \\
&g^{2l}_2(x,y)=  f_{b2}(y)+H(x,y)-\int _x^1g^{2l-1}_1g^{2l}_2(X,y)dX,\quad l\in \N ^*,\nonumber \\
&g^{2l+1}_2(x,y)= f_{b2}(y)+H(x,y)-\int _x^1g^{2l}_1g^{2l+1}_2(X,y)dX,\quad \l \in \N ,\quad (x,y)\in [0,1] \times \omega _\eta ^c.
\end{align}
By induction on $l$ it holds that
\begin{align}\label{approx-scheme-delta-3}
&0\leq g^{2l}_1\leq g^{2l+2}_1\leq g^{2l+3}_1\leq g^{2l+1}_1,\nonumber \\
&0\leq g^{2l}_2\leq g^{2l+2}_2\leq g^{2l+3}_2\leq g^{2l+1}_2,\quad l\in \N .
\end{align} 
Moreover,
\begin{align*}
&\int _{[0,1] \times \omega _\eta ^c}g_j^{2l}(x,y)dxdy\leq \int _0^1f_{bj}(y)dy+\int _{[0,1] \times \omega _\eta ^c}(G+H)(x,y)dxdy,\quad j\in \{ 1,2\} ,\quad l\in \N .
\end{align*}
By the monotone convergence theorem, $(g^{2l})_{l\in \N }$ (resp. $(g^{2l+1})_{l\in \N }$) increasingly (resp. decreasingly) converges in $L^1([0,1] \times \omega _\eta ^c)$ and almost everywhere on $[0,1] \times \omega _\eta ^c$ to some $g$ (resp. $h$). By the dominated convergence theorem,
\begin{align*}
&\lim _{l\rightarrow +\infty }g^{2l}_1g^{2l-1}_2= g_1h_2\quad \text{and}\quad \lim _{l\rightarrow +\infty }g^{2l+1}_1g^{2l}_2= h_1g_2\quad \text{in}\quad L^1([0,1] \times \omega _\eta ^c).
\end{align*}
Consequently,
\begin{align*}
&g_1(x,y)= f_{b1}(y)+G(x,y)-\int _0^xg_1h_2(X,y)dX,\\
& h_1(x,y)= f_{b1}(y)+G(x,y)-\int _0^xh_1g_2(X,y)dX,\\
&g_2(x,y)= f_{b2}(y)+H(x,y)-\int _x^1h_1g_2(X,y)dX= g_2(0,y)-G(x,y)+\int _{0}^xh_1g_2(X,y)dX,\\
&h_2(x,y)= f_{b2}(y)+H(x,y)-\int _x^1g_1h_2(X,y)dX= h_2(0,y)-G(x,y)+\int _{0}^xg_1h_2(X,y)dX,\\
&\hspace*{4.6in} (x,y)\in [0,1] \times \omega _\eta ^c,
\end{align*}
and
\begin{align}\label{comparison}
&h_1\geq g_1,\quad h_2\geq g_2,\quad (x,y)\in [0,1] \times \omega _\eta ^c.
\end{align}
Hence
\begin{align*}
&(h_1-g_1)(1,y)= -(h_2-g_2)(0,y),
\end{align*}
so that, by \eqref{comparison},
\begin{align*}
&g_1(1,y)= h_1(1,y),\quad g_2(0,y)= h_2(0,y).
\end{align*}
Consequently, $h_1-g_1= h_2-g_2$, $g_1h_2-h_1g_2= (g_1-g_2)(h_1-g_1)$ and
\begin{align}\label{h1-g1}
&(h_1-g_1)(x,y)= \int _0^x(g_1-g_2)(h_1-g_1)(X,y)dX.
\end{align}
It follows from $(h_1-g_1)(0,y)= 0$ and the boundedness of $(g_1,g_2)$ on $[0,1] \times \omega _\eta^c $
that $h_1-g_1= 0$ and $(g_1, g_2)= (h_1,h_2)$ on $[0,1] \times \omega _\eta^c $. 
Hence the whole sequence $(g_1^l, g_2^l)_{l\in \N }$ converges to $(g_1,g_2)$ in $L^1([0,1]\times \omega_\eta^c)$. Letting $\eta\rightarrow 0$ and using (2.16), the convergence holds in 
$L^1([0,1]^2)$.\\
\\
Given $\bar{\epsilon} >0$, choose $l_0$ big enough so that $\parallel g^{2l_0}_1-g^{2l_0+1}_1\parallel _{L^1}<\bar{\epsilon}  $, then $k_0$ big enough so that
\begin{align*}
&\parallel f^{k, 2l_0+1}_1-g^{2l_0+1}_1\parallel _{L^1}\leq \bar{\epsilon}  \quad \text{and}\quad \parallel  f^{k,2l_0}_1-g^{2l_0}_1\parallel _{L^1}\leq \bar{\epsilon}  ,\quad k\geq k_0.
\end{align*}
Hence $\parallel f_1^{k,2l_0+1}-f_1^{k,2l_0}\parallel_{L^1}\leq 3\bar{\epsilon} $ \hspace*{0.02in}for $k\geq k_0$. Then
\begin{align*}
&\parallel F^k_1-F_1^{k^\prime }\parallel _{L^1}\\
&\leq \parallel F^k_1-F_1^{k^\prime }\parallel _{L^1((\Omega ^{\epsilon \Lambda }_{k1})^c)}+2c_b\epsilon \quad \text{by   }\eqref{small-mass}\\
&\leq \parallel F^k_1-f^{k,2l_0}_1\parallel_{L^1((\Omega ^{\epsilon \Lambda }_{k1})^c)}+\parallel F_1^{k^\prime }-f^{k^\prime ,2l_0}_1\parallel _{L^1((\Omega ^{\epsilon \Lambda }_{k1})^c)}+\parallel f^{k,2l_0}_1-f^{k^\prime ,2l_0}_1\parallel_{L^1} +2c_b\epsilon \\
&\leq \parallel f^{k,2l_0+1}_1-f^{k,2l_0}_1\parallel_{L^1}+\parallel f^{k^\prime ,2l_0+1}_1-f^{k^\prime ,2l_0}_1\parallel _{L^1}+\parallel f^{k,2l_0}_1-f^{k^\prime ,2l_0}_1\parallel_{L^1}+2c_b\epsilon \quad \text{by   }\eqref{approx-scheme-delta-2bis}\\
&\leq 8\bar{\epsilon} +2c_b\epsilon ,\quad k\geq \max \{ k_0, \exp(\frac{3c_b}{\epsilon })\} ,\quad k^\prime \geq \max \{ k_0, \exp(\frac{3c_b}{\epsilon })\} .
\end{align*}
And so $(F_1^k)$ is a Cauchy sequence in $L^1([0,1]^2)$ with the limit equal to the weak limit $F_1$. Similarly, $(F_j^k)_{2\leq j\leq 4}$ is a Cauchy sequence in $(L^1([0,1]^2))^3$ with the limit equal to the weak limit $(F_j)_{2\leq j\leq 4}$. \cqfd
%
%
%
%
\begin{lemma}\label{F-renormalized-solution}
\hspace*{1.in}\\
The limit $F$ of $(F^k)_{k\in \N ^*}$ in $L^1([0,1]^2)$ is a renormalized solution to the Broadwell model \eqref{eq-broadwell}.
\end{lemma}
\underline{Proof of Lemma \ref{F-renormalized-solution}.}\\
Start from a renormalized formulation of \eqref{Fk-1},
\begin{align}\label{renorm-1}
&\int _0^1\varphi_1(1,y)\ln \big( 1+F^k_1(1,y)\big) dy- \int_0^1\varphi _1(0,y)\ln \big( 1+f_{b1}(y)\wedge \frac{k}{2}\big) dy\nonumber\\
&-\int_0^1\int_0^1\ln\big( 1+F^k_1(x,y)\big) \partial_x\varphi _1(x,y)dxdy\nonumber \\
&= \int _0^1\int_0^1\varphi _1(x,y)\frac{F^k_3F^k_4}{(1+F^k_1)(1+\frac{F^k_3}{k})(1+\frac{F^k_4}{k})}(x,y)dxdy\nonumber\\
&-\int _0^1\int _0^1\varphi _1(x,y)\frac{F^k_1F^k_2}{(1+F^k_1)(1+\frac{F^k_1}{k})(1+\frac{F^k_2}{k})}(x,y)dxdy,
\end{align}
for test functions $\varphi\in (C^1([0,1]^2))^4$.
Using the strong $L^1$ convergence of the sequence $(F^k)$ to pass to the limit when $k\rightarrow +\infty $ in the left hand side of \eqref{renorm-1}, gives in the limit,
\begin{align*}
&\int _0^1\varphi_1(1,y)\ln \big( 1+F_1(1,y)\big) dy- \int_0^1\varphi _1(0,y)\ln \big( 1+f_{b1}(y)\big) dy\\
&-\int_0^1\int_0^1\ln \big( 1+F_1(x,y)\big) \partial_x\varphi _1(x,y)dxdy.
\end{align*}
For the passage to the limit when $k\rightarrow +\infty $ in the right hand side of \eqref{renorm-1}, given $\eta >0$ there is a subset $A_\eta $ of $[0,1]^2$ with $\lvert A_\eta ^c\rvert <\eta $, such that  up to a subsequence, $(F^k)_{k\in \N ^*}$ uniformly converges to $F$ on $A_\eta $ and $F\in L^\infty (A_\eta )$. Passing to the limit when $k\rightarrow +\infty $ on $A_\eta $ is straightforward. Moreover,
\begin{align*}
&\lim _{\eta \rightarrow 0}\int _{A_\eta ^c}\varphi \frac{F_1F_2}{1+F_1}(x,y)dxdy= 0\quad \text{and}\quad \lim _{\eta \rightarrow 0}\int _{A_\eta ^c}\varphi \frac{F^k_1F^k_2}{(1+F^k_1)(1+\frac{F^k_1}{k})(1+\frac{F^k_2}{k})}(x,y)dxdy= 0,
\end{align*}
uniformly with respect to $k$, since 
\begin{align*}
&\frac{F_1}{1+F_1}\leq 1,\quad \frac{F^k_1}{(1+F^k_1)(1+\frac{F^k_1}{k})(1+\frac{F^k_2}{k})}\leq 1,\quad \text{and}\quad \lim_{\eta\rightarrow 0}\int_{A^c_\eta}
F^k_2 =0,
\end{align*}
uniformly with respect to $k$.\\
The gain term can be estimated as follows. The uniform boundedness of the entropy production term of $(F^k)$ is given in Lemma \ref{mass-entropy}. A convexity argument together with the $L^1$ convergence of $(F^k)$ to $F$ (see \cite{DPL}), imply that
\begin{align}\label{renorm-2}
&\int (F_1F_2-F_3F_4)\ln \frac{F_1F_2}{F_3F_4}(x,y)dxdy\leq c_b.
\end{align}
It follows that, for any $\gamma >1$, 
\begin{align*}
\int _{A_\eta ^c}\lvert \varphi \rvert \frac{F_3F_4}{1+F_1}(x,y)dxdy&\leq \frac{c}{\ln \gamma }+c\gamma \int _{A_\eta ^c}\frac{F_1F_2}{1+F_1}(x,y)dxdy\\
&\leq \frac{c}{\ln \gamma }+c\gamma \int _{A_\eta ^c}F_2(x,y)dxdy,
\end{align*}
which tends to zero when $\eta \rightarrow 0$. Similarly, using \eqref{production-entropy},
\begin{align*}
&\int_{A^c_\eta} |\varphi |\frac{F^k_3F^k_4}{(1+F^k_1)(1+\frac{F^k_3}{k})(1+\frac{F^k_4}{k})}(x,y)dxdy\\
&\leq c\int _{A^c_\eta}\frac{F^k_3F^k_4}{(1+F^k_1)(1+\frac{F^k_3}{k})(1+\frac{F^k_4}{k})}(x,y)dxdy\\
&\leq \frac{c}{\ln \gamma}+C\gamma \int_{A^c_\eta}
\frac{F^k_1F^k_2}{(1+F^k_1)(1+\frac{F^k_1}{k})(1+\frac{F^k_2}{k})}(x,y)dxdy\\
&\leq \frac{C}{\ln \gamma}+C\gamma \int_{A^c_\eta}
F^k_2(x,y)dxdy,
\end {align*}
which tends to zero when $\eta\rightarrow 0$, uniformly in $k$. It follows that the right hand side of \eqref{renorm-1} converges to 
\begin{eqnarray*}
\int _0^1\int_0^1\varphi(x,y)\frac{F_3F_4}{1+F_1}(x,y)dxdy
-\int _0^1\varphi(x,y)\frac{F_1F_2}{1+F_1}(x,y)dxdy,
\end{eqnarray*}
when $k\rightarrow +\infty $. Consequently, $F_1$ satisfies the first equation of \eqref{eq-broadwell} in renormalized form. 
It can be similarly proven that $(F_j)_{2\leq j\leq 4}$ is solution to the last equations of \eqref{eq-broadwell}.
\cqfd
This completes the proof of Theorem 2.1.


\begin{thebibliography}{99}
\bibitem
{AN} L. Arkeryd, A. Nouri, {\it On the stationary Povzner equation in $\R^n$}  J. Math. Kyoto Univ. 39 (1) (1999), 115-153.
%
\bibitem
{AN1} L. Arkeryd, A. Nouri, {\it The stationary Boltzmann equation in the slab with given mass for hard and soft forces}, Annali della Scuole Normale Superiore di Pisa Cl. Sci.Mat., Ser IV, Vol XXXVII (1998), 533-566.
%
%
\bibitem
{B} A. Bobylev, {\it Exact solutions of discrete kinetic models and stationary problems for the plane Broadwell model}, Math. Meth. Appl. Sci. (4) 19 (1996), 825-845.
%
\bibitem
{BT} A. Bobylev, G. Toscani, {\it Two dimensional alf-space problems for the Broadwell discrete velocitry model}, Cont. Mechanics and Thermodyamics 8 (1996), 257-274.
%
\bibitem
{BV} A. Bobylev, M. Vinerean-Bernhoff, {\it Discrete velocity models of the Boltzmann equation and conservation laws}, Kinetic and Related Models 3 (2008), 335-358.
%
\bibitem{CIS} C. Cercignani, R. Illner, M. Shinbrot, {\it A boundary value problem for the 2-dimensional Broadwell model}, Comm. Math. Phys. 114 (1988), 687-698.
%
\bibitem{DPL} R. J. DiPerna,  P. L. Lions {\it On the Cauchy problem for Boltzmann equations: Global existence and weak stability}, Ann. Math 130 (1989), 321-366.
%
\bibitem{IP} R. Illner, T. Platkowski, {\it Discrete velocity models of the Boltzmann equation: survey on the mathematical aspects of the theory}, Siam Rev. 30 (1988), 213-255.
%
\bibitem
{Ily} O. V. Ilyin, {\it Symmetries, the current function, and exact solutions for Broadwell's two-dimensional stationary kinetic model}, Teor. Mat. Fiz. 179 (2014), 350-359.
%
\bibitem{K}
A. N. Kolmogorov, {\it \"Uber Kompaktheit der Funktionenmengen bei der Konvergenz im Mittel},
Nachr. Ges. Wiss. G\"ottingen 9 (1931), 60-63.
%
\bibitem{R}
M. Riesz, {\it Sur les ensembles compacts de fonctions sommables}, Acta Szeged Sect. Math. 6 (1933), 136-142.

\end{thebibliography}
\end{document}